\documentclass[twocolumn,amsmath,amssymb,aps,prc,floatfix]{revtex4-1}
\usepackage[english]{babel}
\usepackage{amsfonts}
\usepackage{graphicx}
\usepackage{color}
\usepackage{subfigure}
\usepackage{hyperref}
\usepackage{setspace}    
\usepackage{multirow}
\graphicspath{{./figures/}}
\begin{document}

\title{Baryon-antibaryon dynamics in relativistic heavy-ion collisions}
\author{E. Seifert}
\affiliation{Institut f\"{u}r Theoretische Physik, Universit\"{a}t Gie\ss en, Germany}

\author{W. Cassing}
\affiliation{Institut f\"{u}r Theoretische Physik, Universit\"{a}t Gie\ss en, Germany}
    \date{\today}
    \begin{abstract}
The dynamics of baryon-antibaryon annihilation and reproduction ($B
{\bar B} \leftrightarrow 3 M$)  is studied within the
Parton-Hadron-String Dynamics (PHSD) transport approach for Pb+Pb
and Au+Au collisions  as a function of centrality from lower Super Proton
Synchrotron (SPS) up to Large Hadron Collider (LHC) energies on the
basis of the quark rearrangement model (QRM). At Relativistic
Heavy-Ion Collider (RHIC) energies we find a small net reduction of
baryon-antibaryon ($B {\bar B}$) pairs while for the LHC energy of
$\sqrt{s_{NN}}$ = 2.76 TeV a small net enhancement is found relative to
calculations without annihilation (and reproduction) channels.
Accordingly, the sizeable difference between data and statistical
calculations in Pb+Pb collisions at $\sqrt{s_{NN}}$= 2.76 TeV for
proton and antiproton yields \cite{53}, where a deviation of 2.7
$\sigma$ was claimed by the ALICE Collaboration, should not be
attributed to a net antiproton annihilation. This is in line with
the observation that no substantial deviation between the data and
statistical hadronization model (SHM) calculations is seen for
antihyperons, since according to the PHSD analysis the antihyperons
should be modified by the same amount as antiprotons. As the PHSD
results for particle ratios are in line with the ALICE data (within
error bars) this might point towards a deviation from statistical
equilibrium in the hadronization (at least for protons/antiprotons).
Furthermore, we find that the $B {\bar B} \leftrightarrow 3 M$
reactions are more effective at lower SPS  energies where a net
suppression for antiprotons and antihyperons up to a factor of 2 --
2.5 can be extracted from the PHSD calculations for central Au+Au collisions.

        \par
        PACS: 24.10.-i; 24.10.Cn; 24.10.Jv; 25.75.-q; 14.65.-q
    \end{abstract}
    \maketitle

\section{Introduction}
Relativistic and ultra-relativistic heavy-ion collisions offer the
unique possibility to study a new phase of matter, i.e. a
quark-gluon plasma (QGP), as well as possibly the phase boundary
between the hadronic and partonic phase.  Lattice
Quantum-Chromo-Dynamics (lQCD) calculations suggest that at
vanishing baryon chemical potential ($\mu_B$=0) there is a crossover
phase transition from hadronic to partonic degrees of freedom
\cite{lQCD,lqcd0,LQCDx,Peter,Lat1,Lat2} for the deconfinement phase
transition as well as for the restoration of chiral symmetry.
However, at some finite baryon chemical potential the crossover
might turn  to  a first-order phase transition implying a critical
endpoint in the QCD phase diagram \cite{CBMbook}. Since lattice
calculations so far suffer from the fermion-sign problem, no first
principle information on the phase boundary can be extracted from
lQCD at large $\mu_B$, whereas at low $\mu_B$ Taylor expansions of
the thermodynamic potential (in powers of $\mu_B/T$) provide an
alternative solution. The studies in Refs. \cite{Karsch17,Zoltan17}
show that for heavy-ion reactions at Relativistic Heavy-Ion Collider
(RHIC) and Large Hadron Collider (LHC) energies the phase boundary
is a crossover and the critical temperature for deconfinement $T_c$
is practically the same as at $\mu_B$=0.

Due to the high energy densities reached in Au+Au (Pb+Pb) collisions
at RHIC and LHC energies as well as strong partonic interactions the
final hadron yields turn out experimentally to be close to thermal
and chemical equilibrium as described by a grand-canonical ensemble
of non-interacting hadronic states (with excluded volume
corrections) \cite{STM0,STM01,STM02,STM03,STM04,STM05,STM2,STM1}. In
fact, the thermal analysis of hadron yields at midrapidity show a
high degree of thermalization \cite{PBM17}, however, a sizeable
difference between data and statistical calculations pop up in Pb+Pb
collisions at $\sqrt{s_{NN}}$= 2.76 TeV for proton and antiproton
yields \cite{53}, where a deviation of 2.7 $\sigma$ is obtained
\cite{PBM17}. It has been argued in Refs. \cite{53,Stock} that this
deviation might be due to final state hadronic $B {\bar B}$
annihilation after chemical freezeout. On the other hand such a
reduction was not seen in the relative yields of strange baryons/antibaryons  to
pions \cite{62}. In Refs. \cite{Stock} the UrQMD transport model
\cite{Bass,Bleicher} has been employed as an hadronic `afterburner'
to evaluate the final-state interactions and in particular the
effects from $B {\bar B}$ annihilation after chemical freezeout,
however, the backward channels had been discarded thus violating
detailed balance \cite{Francesco}. This issue has been further
addressed in Ref. \cite{Pratt} in a simplified model for the
space-time evolution but incorporating detailed balance for the
chemical reactions. In the latter study it was found that a net
$B{\bar B}$ reduction by annihilation in central Pb+Pb collisions at
$\sqrt{s_{NN}}$ = 2.76 GeV of $\sim$ 40\% might result thus coping
approximately with the experimental observation in Ref. \cite{53}. A more refined
approach - incorporating detailed balance - has been proposed in
Ref. \cite{Huovinen} which solves chemical rate equations on top of
2+1 hydrodynamic evolution. At RHIC energies the authors report a
reduction of $B{\bar B}$ pairs by about 15-20 \%; results for LHC
energies from this model are not known to the authors. Nevertheless,
the impact of $B {\bar B}$ annihilation and reproduction by the
inverse many-body channels should be calculated on a fully
microscopic basis including detailed balance.

A first step in this direction has been taken in Ref.
\cite{Cassing:2001ds} where the three-body fusion of nonstrange
pseudoscalar and vector mesons to $B {\bar B}$ pairs has been
incorporated in the Hadron-String Dynamics (HSD) transport approach
\cite{HSD} that preferentially describes the hadronic phase and
provides results close to the UrQMD transport model
\cite{Bass,Bleicher} for SPS energies as demonstrated in Refs.
\cite{Weber,Brat04}. In Ref. \cite{Cassing:2001ds} the matrix
element squared for baryon-antibaryon annihilation
has been extracted from the experimental data on $p
{\bar p}$ annihilation and the three-body meson channels have been
determined on the basis of detailed balance. It was found that in
central collisions of heavy nuclei at SPS energies the annihilation
of antinucleons is almost compensated by the inverse recreation
channels. A recent extension of the model has been presented in Ref.
\cite{Paper1} within the Parton-Hadron-String Dynamics (PHSD) approach
\cite{PRC08,PHSD,Bratkovskaya:2011wp} where the full strangeness
sector has been included for the $2 \leftrightarrow 3$ reactions.
The resulting model (denoted by PHSD4.0) has been applied to central
Pb+Pb collisions in the SPS energy regime and it was found again
that $B {\bar B}$ annihilation and reproduction compensate each
other to a large extend.

We recall that the PHSD transport approach
\cite{PRC08,PHSD,Bratkovskaya:2011wp} superseeds the HSD approach by
a couple of aspects that become essential with increasing bombarding
energy: \begin{itemize} \item{the formation of an initial partonic
phase with quark and gluon quasiparticle properties that are fitted
to lattice QCD results in thermodynamic equilibrium} \item{A
dynamical hadronization scheme on the basis of covariant transition
rates} \item{Inclusion of further hadronic reactions in the
strangeness sector with full baryon-antibaryon symmetry}
\item{Inclusion of essential aspects of chiral symmetry restoration
in the hadronic phase \cite{Cas16}.} \end{itemize} Whereas the
latter developments are important for the lower SPS energy regime to
account for the strangeness enhancement seen experimentally in
heavy-ion collisions, the formation of a partonic phase is mandatory
to understand the physics at higher SPS, RHIC and LHC energies. This
has been demonstrated in a couple of PHSD studies in the past for
heavy-ion reactions from $\sqrt{s_{NN}}$ = 4 GeV to 2.76 TeV
\cite{AlesPaper,review,Linnyk:2015tha,Konchakovski:2012yg}. Since
multistrange baryons and antibaryons at top SPS energies and above
no longer stem from string fragmentation (as in HSD
\cite{Cassing:2001ds}) but preferentially from hadronization at
energy densities around 0.5 GeV/fm$^3$ the issue of three-meson
fusion reactions for the formation of baryon-antibaryon ($B {\bar
B}$) pairs and the annihilation of  $B {\bar B}$ pairs to multiple
mesons has to be investigated (in addition to Ref. \cite{Paper1}) at
RHIC and LHC energies.

This work is organized as follows: In Sec. 2 we recapitulate shortly
the ingredients of PHSD and the quark rearrangement model for
baryon-antibaryon annihilation and recreation ($B\bar
B\leftrightarrow 3M$) in the version 4.0 \cite{Paper1}.  In Sec. 3
we present results for antibaryons and multi-strange baryons from
PHSD simulations for central Pb+Pb (Au+Au) collisions at RHIC and
LHC energies in comparison to experimental data and then focus on
the centrality dependence of baryons and antibaryons. We will
compare simulations using \begin{itemize} \item{the
baryon-antibaryon annihilation and formation} \item{only ($B {\bar
B}$) annihilation} \item{without the $2 \leftrightarrow 3$
channels.} \end{itemize} Global excitation functions for mesons,
baryons and antibaryons will be provided in Sec. 4 as well as
excitation functions for the impact of final-state interactions and
in particular the $B {\bar B} \leftrightarrow 3 M$ reactions. We
conclude our study with a summary in Sec. 5.

\section{Reminder of the PHSD transport approach}\label{sec:PHSD}
The Parton-Hadron-String Dynamics (PHSD) is a microscopic covariant
transport approach whose formulation is based on the Kadanoff-Baym
equations \cite{Kadanoff-Baym,Schwinger:1960qe,Schwinger-Keldysh2,Botermans:1990qi}
for Green's functions in phase-space representation in first order
gradient expansion beyond the quasiparticle approximation \cite{Cassing:1999wx}.
 The PHSD transport approach describes in a
consistent manner the whole time evolution of a relativistic heavy-ion
collision as it incorporates a hadronic and a partonic phase as well as
dynamical transitions between the respective degrees-of-freedom.
The properties of the quarks, antiquarks and gluons in the QGP phase are
described by the Dynamical Quasi-Particle Model (DQPM) \cite{Peshier:2004bv,
Peshier:2004ya}, whose three parameters are fixed to reproduce the lQCD
equation-of-state at vanishing baryon chemical potential and which is
based on effective propagators for the partons. The quarks and gluons
have finite masses as well as widths that are given, respectively, by the
real and imaginary parts of the retarded self energies resulting from
two-particle-irreducible diagrams of the effective full propagators.

In PHSD simulations of nucleus-nucleus collisions color neutral strings
(described by the FRITIOF Lund model \cite{NilssonAlmqvist:1986rx}) are
formed from the initial hard nucleon-nucleon scatterings. These strings
decay into ``prehadrons'' with a formation time of $\approx$0.8\,fm/c in
which they do not interact. The string ends are identified with ``leading
hadrons'' and may interact instantly with reduced cross sections according
to the constituent quark model \cite{HSD}. In case that the local energy
density surpasses the critical value of $\epsilon_c\approx$0.5\,GeV/fm$^3$
the prehadrons dissolve into colored effective quarks, antiquarks and
gluons given by the DQPM at the given local energy density. These partons
then propagate in their self-generated mean field and interact via quasi
elastic $2\rightarrow 2$ collisions between quarks, antiquarks and gluons.
Additionally, $q\bar q$ pairs may annihilate into a gluon and gluons may
decay to $q\bar q$ pairs.
As the system expands the local energy density will drop close to or
below $\epsilon_c$ and will start to hadronize into off-shell mesons and
baryons. During hadronization the energy, three-momentum, and quantum
numbers are conserved in each event \cite{PRC08}.
In the hadronic phase the particles interact with each other (as in
HSD)  via elastic and inelastic collisions satisfying the detailed
balance relations. The cross sections are taken from experiments or
effective models.

\subsection{Quark rearrangement model for $B+\bar B$ production and
annihilation}\label{sec:QuarkRearrangement} The quark rearrangement
model (in the context of  $B {\bar B}$ annihilation and
reproduction) goes back to Ref. \cite{Cassing:2001ds}  and has been
extended to the SU(3)$_{flavor}$ sector recently \cite{Paper1}. It
is based on the experimental observation of a dominant annihilation
of $p\bar p$ into (on average) 5 pions at invariant energies
$2.3\,\mathrm{GeV}\leq\sqrt s \leq 4\,\mathrm{GeV}$. Now the final
number of 5 pions may be interpreted as an initial annihilation into
$\pi\rho\rho$ with the $\rho$ mesons decaying subsequently into two
pions each. The channel $\pi\pi\rho$ then leads to 4 final pions,
the channel $\pi\omega\rho$ to 6 final pions, the channel
$\rho\omega\rho$ to 7 final pions etc. Accordingly, the
baryon-antibaryon annihilation in the first step is a $2\rightarrow 3$
reaction with a conserved number of quarks and antiquarks. This is
the basic assumption of the quark rearrangement model which is
illustrated in Fig. 2 of Ref. \cite{Paper1}.  By allowing the mesons
$M_i$ to be any member  of the $0^-$ or $1^-$ nonets one can
describe an arbitrary $B \bar B$ annihilation and recreation by
rearranging the quark and antiquark content, where $B$ is a member
of the baryon octet or decuplet.  This approach gives a realistic
description for $p {\bar p}$ annihilation and we assume that for
other baryon-antibaryon pairs than $p\bar p$ a similar annihilation
pattern holds. Since there are no measurements of annihilation cross
sections other than $p\bar n$ and $p\bar p$ this is our best guess
at present.

\subsection{2$ \leftrightarrow$ 3 reactions in kinetic theory}
The treatment of 2$ \leftrightarrow$ 3 reactions can be incorporated
in the collision term in kinetic theory on a channel-by-channel
basis by employing detailed balance as formulated in Refs.
\cite{Cassing:2001ds,Paper1}. The matrix elements squared $|M^c_{B
{\bar B}}(\sqrt{s})|^2$ for a channel $c$, which - apart from
phase-space integrals - determine the $B {\bar B}$ annihilation
rate, are assumed to be the same for all flavor channels $c$. We
recall that more than 2500 individual mass channels are incorporated
and the 3-body phase-space integrals have to be evaluated for each
of these channels as a function of invariant energy $\sqrt{s}$
\cite{Paper1}. The technical solution to this problem and detailed
tests of the algorithm in a finite box with periodic boundary
conditions have been described in Ref. \cite{Paper1}. The matrix
element squared $|M_{B {\bar B}}(\sqrt{s})|^2$ has been extracted
from the experimental data on $p {\bar p}$ annihilation as a
function of the relative momentum and it has been assumed that the
product of relative velocity (in the center-of-mass system)
$v_{rel}$ and annihilation cross section $\sigma_{ann}$ for other
flavor channels is the same as for the channel $p {\bar p}$. In this
work we discard an explicit strangeness suppression factor which was
found in Ref. \cite{Paper1} (Fig. 19) to have a minor impact on the
actual results for relativistic heavy-ion collisions. Within these
assumptions the kinetic approach has no additional free
parameter when incorporating the 2$ \leftrightarrow$ 3 reactions in
addition to the $1 \leftrightarrow 2$ and $2 \leftrightarrow 2$
channels \cite{AlesPaper}. The actual solution of the covariant
transition rates in case of heavy-ion collisions is performed by
Monte-Carlo employing the in-cell method as in Refs.
\cite{Cassing:2001ds,Paper1}.

\section{PHSD simulations for heavy-ion collisions}\label{sec:PHSDsimulation}

\begin{figure}[t]
{\centering
\includegraphics[width= 0.4\textwidth]{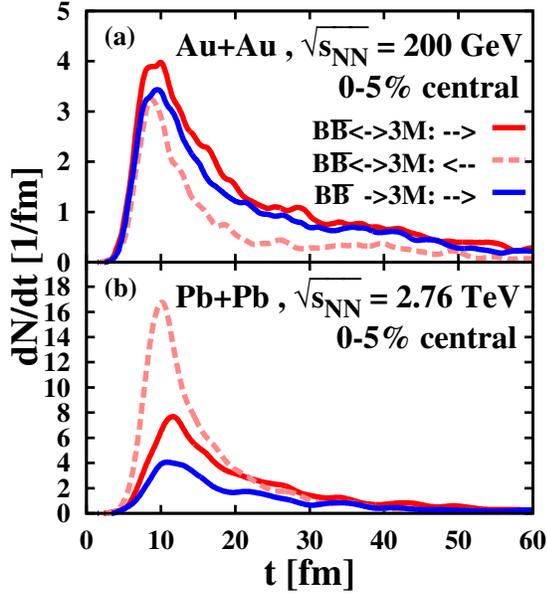}}
\caption{(Color online) The reaction rates of the $B\bar
B\rightarrow 3M$ reactions (solid line) as a function of time in
5\% central Au+Au collisions  at $\sqrt{s_{NN}}$ = 200 GeV a) and Pb+Pb collisions at
$\sqrt{s_{NN}}$ = 2.76 TeV b) integrated over rapidity. The solid blue lines denote
the rates for $B {\bar B}$ annihilation when discarding the
reproduction channels; the red solid lines stand for the $B {\bar
B}$ annihilation rate when including the backward channels whereas
the dashed lines display the reproduction rate in the latter case.
}\label{fig:phsd-rate-t}
\end{figure}

In this section we show the influence of the  $B\bar
B\leftrightarrow 3M$ reactions on heavy-ion collisions in particular
at RHIC and LHC energies in extension of the calculations at SPS
energies in Ref. \cite{Paper1}. Before coming to the actual results
for hadron spectra we compare in Fig. \ref{fig:phsd-rate-t} the
reaction rates for the total baryon-antibaryon annihilation and
formation  from PHSD in 5\% central Pb+Pb (Au+Au) collisions at
$\sqrt{s_{NN}}$ = 200 GeV (a),  and 2.76 TeV (b) integrated over
rapidity. The solid blue lines denote the rates for $B {\bar B}$
annihilation when discarding the reproduction channels; the red
solid lines stand for the $B {\bar B}$ annihilation rate when
including the backward channels whereas the dashed lines display the
reproduction rate in the latter case. The meson-fusion rate
dominates at early times at  the LHC energy over the  $B {\bar B}$
annihilation rate (b) while the situation is inverse at the top RHIC
energy (a). Without regeneration of  $B {\bar B}$ pairs (blue solid
lines) the annihilation rates are lower than in case of $B {\bar B}$
reproduction which is, however, an unphysical limit and displayed
only for orientation. The explicit dependence of ratios versus
$\sqrt{s_{NN}}$ will be discussed in Sec. \ref{sec:ExcitationFunctions}.

\subsection{Hadron transverse-momentum spectra at RHIC and LHC}

\begin{figure}[h]
\centering
\includegraphics[width=0.35\textwidth]{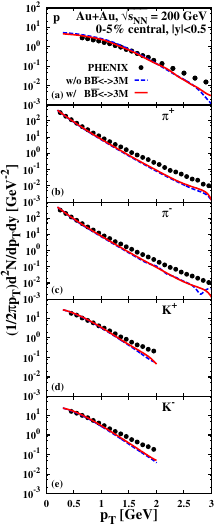}
\caption{(Color online) The transverse momentum spectra for protons,
positive and negative pions as well as for kaons and antikaons from
PHSD at midrapidity in comparison to the data from the PHENIX Collaboration
\cite{Ex1} for 5\% central Au+Au collisions
at $\sqrt{s_{NN}}$=200 GeV. The full red lines show the results of
calculations with the $2 \leftrightarrow 3$ reactions included while
dashed lines correspond to calculations with the $2 \leftrightarrow
3$ reactions discarded. }\label{2}
\end{figure}
\begin{figure}[h]
\centering
\includegraphics[width=0.35\textwidth]{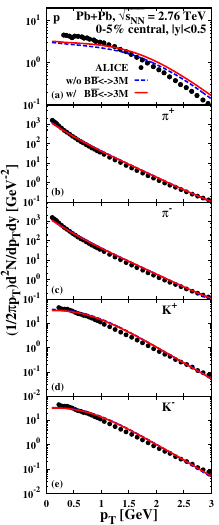}
\caption{(Color online) The transverse momentum spectra for protons,
positive and negative pions as well as for kaons and antikaons from
PHSD at midrapidity in comparison to the data from the ALICE
Collaboration \cite{Ex4,Ex5,Ex6,Ex7} for 5\% central Pb+Pb collisions at
$\sqrt{s_{NN}}$=2.76 TeV. The full red lines show the results of
calculations with the $2 \leftrightarrow 3$ reactions included while
dashed lines correspond to calculations with the $2 \leftrightarrow
3$ reactions discarded.}\label{3}
\end{figure}
\begin{figure}[h]
\centering
\includegraphics[width=0.35\textwidth]{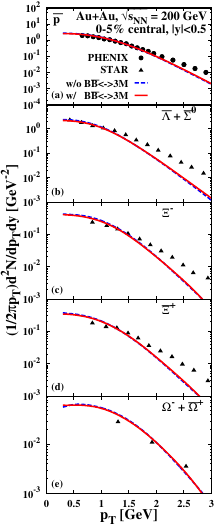}
\caption{(Color online) The transverse momentum spectra for ${\bar
p}, {\bar \Lambda} + {\bar \Sigma}^0, \Xi^-, {\bar \Xi}^+$ and
$\Omega^- + {\bar \Omega}^+$ from PHSD at midrapidity in comparison
to the data from the PHENIX and STAR Collaborations \cite{Ex1,Ex2,Ex3}
for 5\% central Au+Au collisions at $\sqrt{s_{NN}}$=200 GeV. The
full red lines show the results of calculations with the $2
\leftrightarrow 3$ reactions included while dashed lines correspond
to calculations with the $2 \leftrightarrow 3$ reactions discarded.
}\label{4}
\end{figure}
\begin{figure}[h]
\centering
\includegraphics[width=0.35\textwidth]{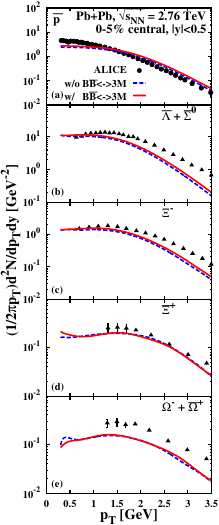}
\caption{(Color online) The transverse momentum spectra for ${\bar
p}, {\bar \Lambda} + {\bar \Sigma}^0, \Xi^-, {\bar \Xi}^+$ and
$\Omega^- + {\bar \Omega}^+$ from PHSD at midrapidity in comparison
to the data from the ALICE Collaboration \cite{Ex5,Ex6} for 5\% central
Pb+Pb collisions at $\sqrt{s_{NN}}$=2.76 TeV. The full red lines
show the results of calculations with the $2 \leftrightarrow 3$
reactions included while dashed lines correspond to calculations
with the $2 \leftrightarrow 3$ reactions discarded. }\label{5}
\end{figure}
\begin{figure}[h]
\centering
\includegraphics[width= 0.35\textwidth]{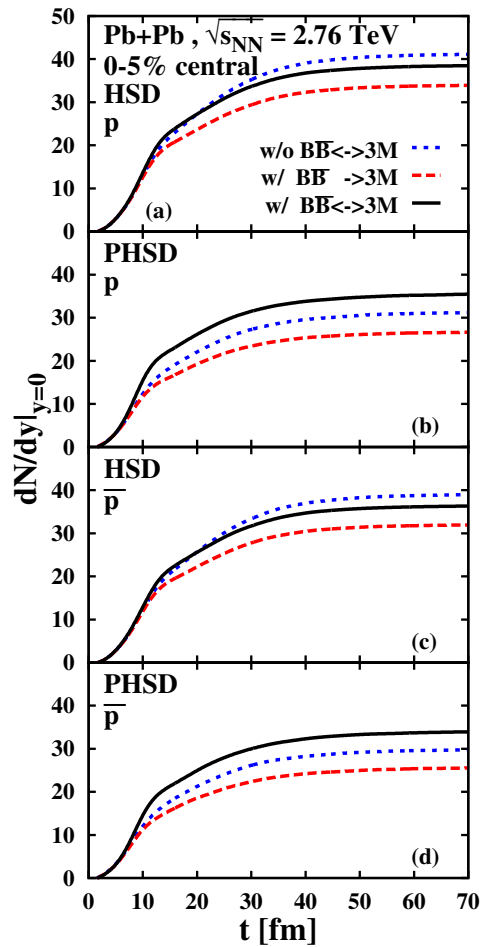}
\caption{Time evolution of the midrapidity yields of the protons (a,b) and
antiprotons (c,d) in 0-5\% central Pb+Pb collisions at $\sqrt{s_{NN}}=2.76$\,TeV
from HSD (a,c) and PHSD (b,d). The dashed red lines
show the results of calculations with only $B {\bar B}$
annihilation, the solid black lines show results with the $2
\leftrightarrow 3$ reactions included while the dotted blue lines
correspond to calculations with the $2 \leftrightarrow 3$ reactions
discarded.}\label{fig:protons-antiprotons}
\end{figure}
\begin{figure}[h]
\centering
\includegraphics[width=0.51\textwidth]{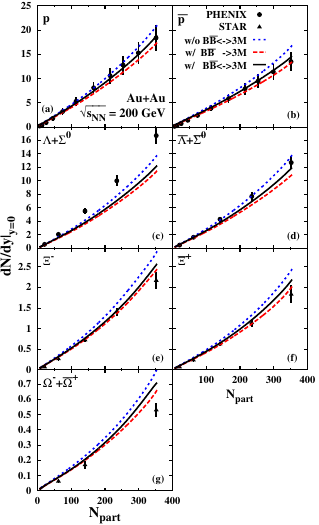}
\caption{(Color online) The rapidity density of baryons and
antibaryons  from PHSD at midrapidity in comparison to  data from
the PHENIX and STAR Collaborations \cite{Ex1,Ex2,Ex3} for 5\% central
Au+Au collisions at $\sqrt{s_{NN}}$=200 GeV. The dashed red lines
show the results of calculations with only $B {\bar B}$
annihilation, the solid black lines show results with the $2
\leftrightarrow 3$ reactions included while dotted blue lines
correspond to calculations with the $2 \leftrightarrow 3$ reactions
discarded. }\label{6}
\end{figure}
\begin{figure}[h]
\centering
\includegraphics[width=0.51\textwidth]{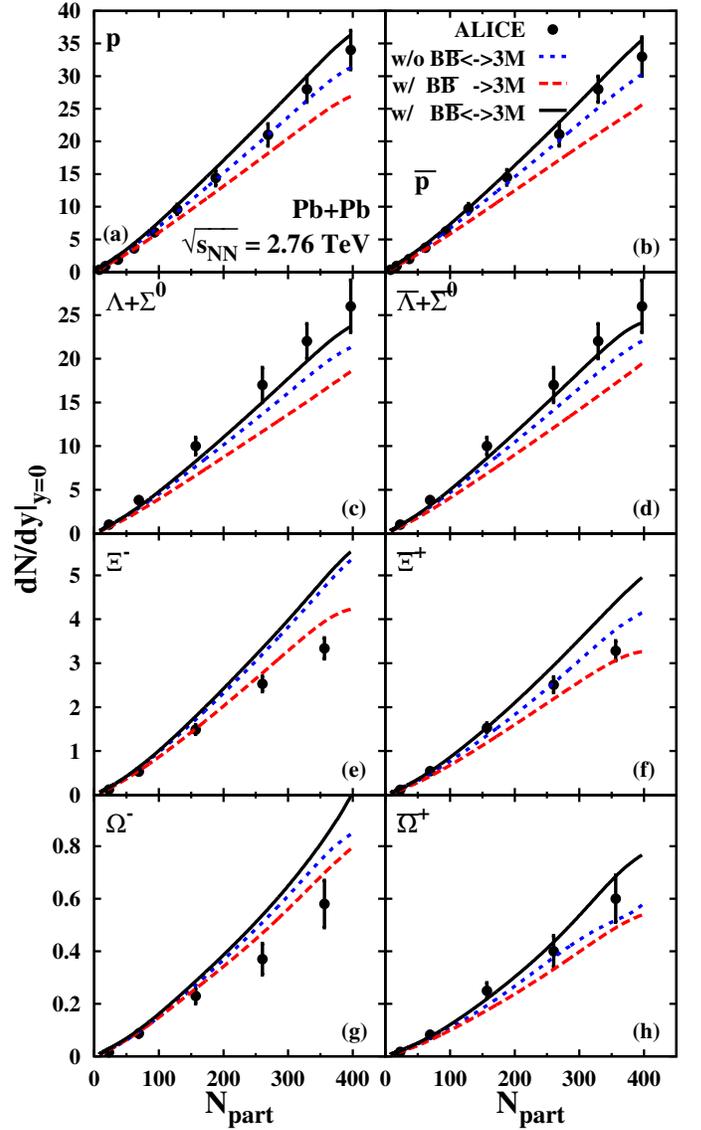}
\caption{(Color online) The rapidity density of baryons and
antibaryons from PHSD at midrapidity in comparison to the data from
the ALICE Collaboration \cite{Ex4,Ex5,Ex6,Ex7} for 5\% central Pb+Pb collisions
at $\sqrt{s_{NN}}$=2.76 TeV. The dashed red lines show the results
of calculations with only $B {\bar B}$ annihilation, the solid black
lines show results with the $2 \leftrightarrow 3$ reactions included
while dotted blue lines correspond to calculations with the $2
\leftrightarrow 3$ reactions discarded. }\label{7}
\end{figure}
\begin{figure}[h]
\centering
\includegraphics[width=0.43\textwidth]{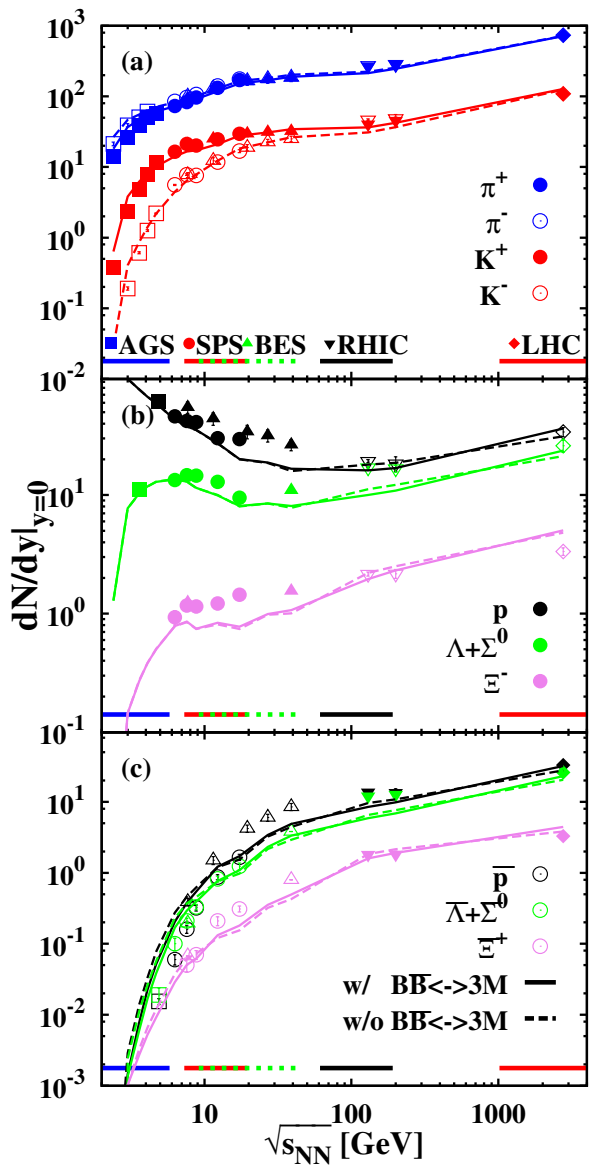}
\caption{(Color online) The midrapidity yields of mesons (a),
baryons (b) and antibaryons (c) from PHSD as a function of the
invariant energy $\sqrt{s_{NN}}$ for central heavy-ion collisions in
comparison with the experimental data taken from Refs. \cite{Ex1,Ex2,
Ex3,Ex4,Ex5,Ex6,Ex7,Ex8,Ex9,Ex10,Ex11,Ex12,Ex13,Ex14,Ex15,Ex16,Ex17,
Ex18,Ex19,Ex20}.
The solid lines refer to calculations including the $B{\bar B}
\leftrightarrow 3 M$ channels while the dashed lines display
calculations without these channels. The particle yields  from PHSD
are connected by lines to draw the eye although experimental data
and calculations do not always correspond to the same centrality
selection (and system) for different bombarding energies. }\label{8}
\end{figure}
\begin{figure}[h]
\centering
\includegraphics[width= 0.33\textwidth]{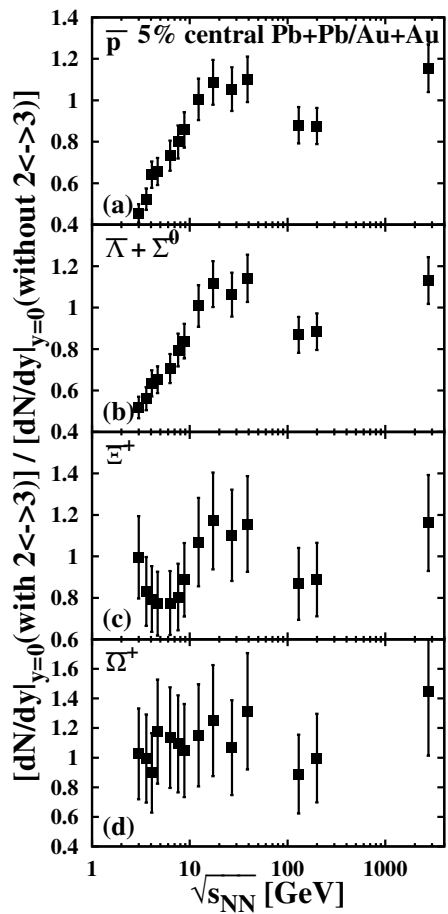}
\caption{(Color online) Ratios of 0-5\% central midrapidity yields
from calculations with the full $B\bar B \leftrightarrow 3M$
reactions to calculations without them for the antibaryons as a
function of the invariant energy $\sqrt{s_{NN}}$.}\label{9}
\end{figure}
\begin{figure}[h]
\centering
\includegraphics[width=0.33\textwidth]{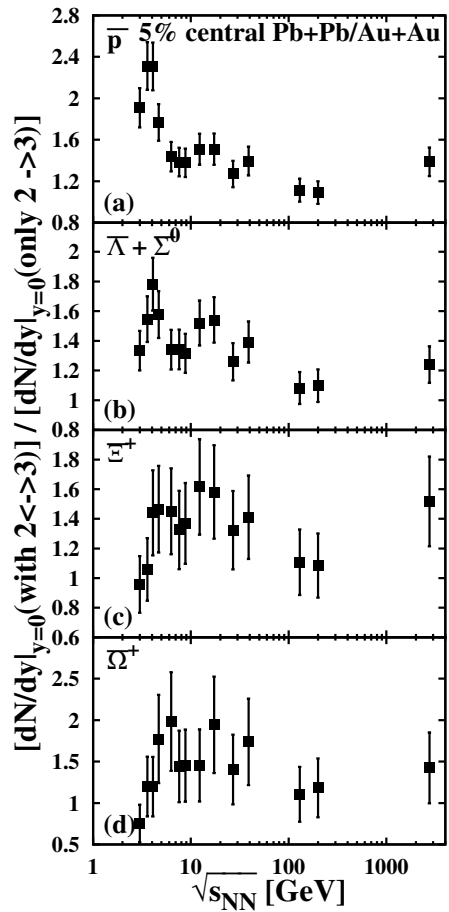}
\caption{(Color online) Ratios of 0-5\% central midrapidity yields
from PHSD calculations with the full $B\bar B \leftrightarrow 3M$
reactions to calculations with only annihilation as a function of
the invariant energy $\sqrt{s_{NN}}$.}\label{10}
\end{figure}

We recall that rapidity and transverse momentum spectra of
antibaryons from PHSD in central Pb+Pb collisions at SPS energies
have been shown in Ref. \cite{Paper1} in comparison to the available
data. We here continue with PHSD results for antibaryons and mesons
in 5\% central Pb+Pb (Au+Au) collisions at the top RHIC energy
($\sqrt{s_{NN}}$=200 GeV) and the LHC energy of $\sqrt{s_{NN}}$=
2.76 TeV. In Fig. \ref{2} we display the calculated transverse
momentum spectra for protons, positive and negative pions as well as
for kaons and antikaons in comparison to the data from the PHENIX
 Collaboration \cite{Ex1}. Whereas the hadron spectra
are quite well described at lower transverse momenta there is a
deficit at high $p_T$ for all hadron species in the PHSD
calculations. We note that the hadron formation at the top RHIC
energy at midrapidity proceeds essentially by hadronization, i.e. by
dynamical coalescence, which implies that the quarks and antiquarks
at hadronization have  softer transverse momenta in PHSD than in
`experiment'. The total hadron densities at midrapidity are only
marginally affected by the underestimated high $p_T$ tail and we may
conclude that the hadron densities within PHSD are sufficiently
realistic such that rather solid results for the annihilation and
fusion rates should emerge. The full red lines show the spectra from
calculations with the $2 \leftrightarrow 3$ reactions included while
the dashed lines correspond to calculations with the $2
\leftrightarrow 3$ reactions discarded. Since there are almost no
differences between the lines we can conclude again that the $2
\leftrightarrow 3$ reactions have practically no impact on baryon
and meson spectra (cf. Ref. \cite{Paper1} for SPS energies).

In Fig. \ref{3} we show the same hadron $p_T$ spectra at midrapidity
for 5\% central Pb+Pb collisions at $\sqrt{s_{NN}}$=2.76 TeV in
comparison to the data from the ALICE Collaboration \cite{Ex4,Ex5,Ex6,Ex7}.
In this case the description of the data is rather good (except for
protons) and again there is no visible impact of the $2
\leftrightarrow 3$ reactions on these transverse momentum spectra.
We note in passing that the flow coefficients $v_n$ (for
$n=2,3,4,5$) from PHSD for this system are also in a very good
agreement with the experimental measurements as shown in Ref.
\cite{Koncha2015}. Thus we may state that the densities of the most
abundant hadrons appear to be  well under control in PHSD in
particular at the LHC energy.

We continue with the antibaryon transverse momentum spectra at
midrapidity for top RHIC and LHC energies, which are displayed in
Figs. \ref{4} and \ref{5}, respectively, in comparison with the data
from the PHENIX, STAR and ALICE Collaborations \cite{Ex1,Ex2,Ex3,Ex5,Ex6}.
Here again the low momentum spectra for ${\bar p}, {\bar \Lambda} +
{\bar \Sigma}^0, \Xi^-, {\bar \Xi}^+$ and $\Omega^- + {\bar
\Omega}^+$ are roughly described at low momenta, however, the high
$p_T$ tails are missed considerably at $\sqrt{s_{NN}}$ = 200 GeV in
Fig. \ref{4} while they look somewhat better at the LHC energy in
Fig. \ref{5}. We note that again there is no visible impact of the
$2 \leftrightarrow 3$ reactions on these transverse momentum spectra.

\subsection{Time evolution of the antiproton and proton yield at midrapidity in different scenarios}
Some further information on the role of the  $2 \leftrightarrow 3$ reactions can be extracted from the actual time evolution of the baryon and antibaryon yields.  
In Fig. \ref{fig:protons-antiprotons} we display the number of formed protons and antiprotons (for $|y|<$ 0.5) as a function of time for a central Pb+Pb collision at $\sqrt{s_{NN}} = 2.76$\,TeV in the following scenarios:
\begin{itemize}
\item{HSD calculation without any annihilation and recreation channels ($B {\bar B} \leftrightarrow 3M$)}
\item{HSD calculation with only the annihilation ($B {\bar B} \rightarrow 3M$) channels}
\item{HSD calculation with all $B {\bar B} \leftrightarrow 3M$ channels included}
\item{PHSD calculation without the $B {\bar B} \leftrightarrow 3M$ channels}
\item{PHSD calculation with only the $B {\bar B} \rightarrow 3M$ channels}
\item{PHSD calculation with all $B {\bar B} \leftrightarrow 3M$ channels included}
\end{itemize}
Whereas the first 3 scenarios do not incorporate any partonic phase the last 3 scenarios do such that a direct
comparison allows to study the relative impact of the QGP phase and the role of the $2 \leftrightarrow 3$ reactions. We find that the time evolution of the protons
is very similar to that of the antiprotons in all scenarios considered. This is essentially due to the fact that at this energy (and midrapidity) the baryon chemical potential is approximately zero  and particle/antiparticle reactions are treated on the same footing in HSD/PHSD. We recall that in HSD `formed' hadrons  only appear for
energy densities of $\epsilon<0.5$\,GeV/fm$^3$ as well as hadronic scatterings and that the production of midrapidity particles is dominated by
\texttt{PYTHIA6.4}, whereas in PHSD (especially at $\sqrt{s_{NN}}=2.76$\,TeV) the midrapidity particles are produced
by hadronization at energy densities $\epsilon \approx 0.5$\,GeV/fm$^3$. Thus formed hadrons in both models appear at roughly the same time but their production mechanism is different and the particles from hadronization in PHSD carry the collective flow from the interacting partonic phase. In general, the yield of protons and antiprotons is higher from  \texttt{PYTHIA6.4} (in HSD) than that from hadronization (in PHSD) by almost 30 \% as seen from the ratio of the blue dotted lines in Fig. \ref{fig:protons-antiprotons}, where the $2 \leftrightarrow 3$ reactions are discarded.
On the other hand, from PHSD we get about 10 \% more pions than from HSD at midrapidity in this limit such that a higher three-meson fusion rate can be expected in PHSD and a higher annihilation rate in HSD.  
When switching on the annihilation channels (dashed red lines) a net proton and antiproton reduction of about 20 \% shows up for HSD and roughly 18 \% for PHSD. However, when accounting for all  $B\bar B\leftrightarrow 3M$ channels (solid black lines) the proton and antiproton abundances are larger than those without the $2 \leftrightarrow 3$ reactions in case of PHSD ( by 12 \%) while for HSD the full calculations still show a tiny net annihilation (by 7 \%). The final proton and antiproton midrapidity yields - with the $2 \leftrightarrow 3$ reactions included - are rather close for HSD and PHSD and differ only by $\sim$ 6 \% which demonstrates that the  $2 \leftrightarrow 3$ channels wash out the memory from the initial production to a large extent.

\subsection{Centrality dependence of baryons and antibaryons at RHIC and LHC}
We continue with $p_T$ integrated rapidity
densities for baryons and antibaryons as a function of centrality in
terms of the number of participating nucleons $N_{part}$ which is
calculated within PHSD. Fig. \ref{6} shows the rapidity density of
baryons and antibaryons  from PHSD at midrapidity in comparison to
data from the PHENIX and STAR Collaborations \cite{Ex1,Ex2,Ex3} for 5\%
central Au+Au collisions at $\sqrt{s_{NN}}$=200 GeV. When discarding
the  $2 \leftrightarrow 3$ reactions (blue dotted lines) the
experimental data are slightly overestimated (except for $\Lambda +
\Sigma^0$), while calculations with only $B {\bar B}$ annihilation
(dashed red lines) show a slight tendency to underestimate the data.
The results from PHSD calculations with the  $2 \leftrightarrow 3$
reactions included are displayed by the black solid lines and lie in
between the other limits. This points towards a small net $B{\bar
B}$ annihilation at the top RHIC energy for all baryons/antibaryons
considered. We will quantify this net annihilation in Sec. \ref{sec:ExcitationFunctions}.

The situation is somewhat different at LHC energies. Fig. \ref{7}
shows the rapidity density of baryons and antibaryons  from PHSD at
midrapidity in comparison to data from the ALICE  Collaboration
\cite{Ex4,Ex5,Ex6,Ex7} for 5\% central Pb+Pb collisions at $\sqrt{s_{NN}}$=2.76
TeV. The blue dotted lines display the calculated results when
discarding the $2 \leftrightarrow 3$ reactions and the dashed red
lines correspond to calculations with only $B {\bar B}$
annihilation.  The results from PHSD calculations with the $2
\leftrightarrow 3$ reactions included are displayed by the black
solid lines and lie in all cases slightly above the other limits
indicating a net $B {\bar B}$ production at the LHC instead of an
absorption. The calculations with only $B {\bar B}$ annihilation
(red dashed line) underestimate the experimental data (except for
$\Xi^-$ and $\Omega^-$). In particular the $p, {\bar p},
\Lambda+\Sigma^0$, and ${\bar \Lambda} + {\bar \Sigma}^0$
multiplicities are (within error bars) in line with experimental
observation at all centralities (when including the $B{\bar B}
\leftrightarrow 3 M$ channels) contrary to the results of the SHM
quoted in Ref. \cite{53}. On the other hand the $\Xi^-, {\bar \Xi}^+,
\Omega^-$ and ${\bar \Omega}^+$ baryons are slightly overestimated in
more central collisions when including the $B{\bar B}
\leftrightarrow 3 M$ channels. We attribute these results to a
deviation from statistical equilibrium in the hadronization
incorporated in PHSD.

\section{Excitation functions}\label{sec:ExcitationFunctions}
In this Sec. we will quantify the net effect of the $B{\bar B}
\leftrightarrow 3 M$ channels for central Pb+Pb (Au+Au) collisions
as a function of the bombarding energy or $\sqrt{s_{NN}}$,
respectively, including the previous results from Ref.
\cite{Paper1}.

\subsection{Hadron yields at midrapidity}
In Fig. \ref{8} we first show the performance of PHSD4.0 with
respect to hadron production (at midrapidity)  in central Pb+Pb
(Au+Au) collisions from  $\sqrt{s_{NN}}$= 3.5 GeV to 2.76 TeV, i.e.
by roughly 3 orders of magnitude in invariant energy. The solid
lines refer to calculations including the $B{\bar B} \leftrightarrow
3 M$ channels while the dashed lines display calculations without
these channels. The particle yields at midrapidity from PHSD are
connected by lines (to draw the eye) although experimental data (taken
from Refs. \cite{Ex1,Ex2,Ex3,Ex4,Ex5,Ex6,Ex7,Ex8,Ex9,Ex10,Ex11,Ex12,
Ex13,Ex14,Ex15,Ex16,Ex17,Ex18,Ex19,Ex20}) and calculations do not always
correspond to the same centrality selection (and system) for different
bombarding energies. However, for given $\sqrt{s_{NN}}$ data and calculation
correspond to the same centrality and collision system. From Fig.
\ref{8} a) we see that PHSD essentially reproduces the experimental
observations for pions, kaons and antikaons in the whole energy
range. We recall that at AGS and SPS energies this is essentially
due to the incorporation of chiral symmetry restoration (cf. Refs.
\cite{AlesPaper,Cas16}). The same holds true for the baryon and
antibaryon excitation functions except for the energy regime 20 GeV
$ < \sqrt{s_{NN}} < $100 GeV where PHSD underestimates the baryons
and antibaryons. The reason for this discrepancy is presently not
understood. However, by comparing the hadron yields from
calculations with (solid lines) and without (dashed lines) the
$B{\bar B} \leftrightarrow 3 M$ channels we find no essential
differences by eye.

\subsection{Quantitative impact of many-body reactions}
In this Subsec. we will quantify the effect of the  $B{\bar B}
\leftrightarrow 3 M$ channels and  $B{\bar B} \rightarrow 3 M$
channels in 5\% central Pb+Pb collisions for 3.5 GeV $ \leq
\sqrt{s_{NN}} \leq$ 2.76 TeV. To this end we show in Fig. \ref{9}
the ratio of the antibaryons ${\bar p}, {\bar \lambda}+{\bar
\Sigma}^0, {\bar \Xi}$ and ${\bar  \Omega}$ (at midrapidity) from
PHSD calculations including the $B{\bar B} \leftrightarrow 3 M$
channels to calculations without them. At low $\sqrt{s_{NN}}
\approx$ 3.5\,GeV we observe a sizeable net annihilation of antiprotons
and antihyperons by about a factor of two which is essentially due
to the fact that here the nucleon density is very large compared to
the antinucleon density. Practically the same holds for the
strangeness $S = \pm$ 1 sector while the net suppression of ${\bar
\Xi}^+$ is only 20\%. For ${\bar \Omega}^+$'s there is no net
suppression within error bars which results from the statistical
errors of both calculations. With increasing invariant energy the
net annihilation of antiprotons and antihyperons disappears at
$\sqrt{s_{NN}} \approx$ 10 GeV, i.e. at the top SPS and lower RHIC
energies. For $\sqrt{s_{NN}}$ = 130 GeV and 200 GeV we find a small
net annihilation for ${\bar p}, {\bar \Lambda}+{\bar \Sigma}^0$, and
$ {\bar \Xi}^+$ which turns to a small enhancement at the LHC energy
of $\sqrt{s_{NN}}$ = 2.76 TeV as quoted before. This is in contrast
to the results of the model calculations in Refs.
\cite{Stock,Pratt}. The small net suppression of antiprotons at the
top RHIC energy, however, is in line with the results from Ref.
\cite{Huovinen} which also incorporate detailed balance for the
annihilation channels. We interpret the tiny enhancement of
antibaryons at the LHC energy to result from the huge meson
abundances which in phase space are slightly overpopulated in PHSD relative
to baryon-antibaryon pairs at hadronization.

In order to investigate the effect of the $B{\bar B}$ annihilation
channels we show in Fig. \ref{10} the ratio of the antibaryons
${\bar p}, {\bar \Lambda}+{\bar \Sigma}^0, {\bar \Xi}^+$ and ${\bar
\Omega}^+$ (at midrapidity) from PHSD calculations including the
$B{\bar B} \leftrightarrow 3 M$ channels to calculations with only
the annihilation channels for the same reactions as in Fig. \ref{9}.
Although this ratio is an unphysical quantity it allows to shed
light on the relative importance of the annihilation channels. For
all antibaryons in Fig. \ref{10} this ratio is larger than unity
which implies that the back-reactions have some impact on the final
antibaryon multiplicities. This effect is most pronounced at lower
SPS energies, where the baryon densities are large compared to the
antibaryon densities, and drops below 50\% enhancement for invariant
energies above about $\sqrt{s_{NN}}=10$\,GeV (within error bars). At top RHIC and LHC
energies these modifications are below the 20\% level since baryon
and antibaryon densities are comparable and all elastic and
inelastic $2 \leftrightarrow 2$ channels are equal for time reversed
states. Only the relative weight of baryons to mesons changes
slightly resulting in ratios greater than unity.

\section{Summary}\label{sec:summary}
In this study we have employed the extended  quark rearrangement
model (QRM) for baryon-antibaryon annihilation ($B {\bar B}
\leftrightarrow 3 M$) from Ref. \cite{Paper1} - incorporated in
PHSD4.0 - for the hadron production in heavy-ion collisions at
ultra-relativistic energies. We recall - using simulations in a box
with periodic boundary conditions - that the numerical
implementation of the quark rearrangement model including the
strangeness sector satisfies the detailed balance $2 \leftrightarrow
3$ relations on a channel-by-channel basis as well as differentially
as a function of the invariant energy $\sqrt{s}$  \cite{Paper1}. It
has been found that the effects from the ($B {\bar B}
\leftrightarrow 3 M$) reaction channels on the meson, baryon and
antibaryon spectra is  only moderate, although nonzero. At the top
RHIC energy we find a small net suppression of $B {\bar B}$ pairs
relative to calculations without these channels whereas at the LHC
energy of $\sqrt{s_{NN}}$ = 2.76 TeV there is even a net enhancement
of $B {\bar B}$ pairs which we attribute to the higher meson
densities. The PHSD net antibaryon enhancement is in contrast to the
results of the model calculations in Refs. \cite{Stock,Pratt} at the
LHC energy whereas the small net suppression of antiprotons at the
top RHIC energy is in line with the results from Ref.
\cite{Huovinen} which also incorporate detailed balance for the
annihilation channels. Accordingly, the sizeable difference between
data and statistical calculations in Pb+Pb collisions at
$\sqrt{s_{NN}}$= 2.76 TeV for proton and antiproton yields
\cite{53}, where a deviation of 2.7 $\sigma$ was obtained
\cite{PBM17}, should not be attributed to a net antiproton
annihilation. On the other hand, no substantial deviation between
the data and SHM calculations is seen for antihyperons \cite{62},
which according to the PHSD analysis should be modified by the same
amount as antiprotons (cf. Fig. \ref{9}).

\begin{figure}[hbt]
\centering
\includegraphics[width= 0.47\textwidth]{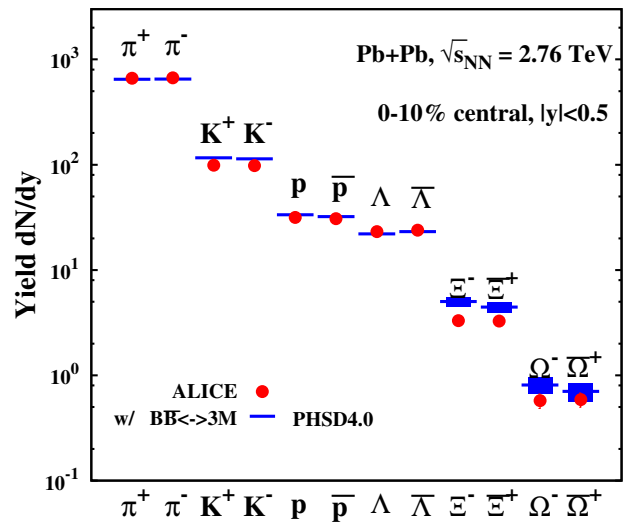}
\caption{(Color online) The midrapidity yields from PHSD4.0
including the $B\bar B\leftrightarrow 3M$ reactions (solid lines)
for 0-10\% central Pb+Pb collisions at $\sqrt{s_{NN}}=2.76$\,TeV.
The data points are taken from Ref. \cite{PBM17}.}\label{11}
\end{figure}

To summarize our results
for the LHC energy we show in Fig. \ref{11} the particle ratios at
midrapidity from 10\% central Pb+Pb collisions in comparison to the
data from the ALICE Collaboration. Since the PHSD results are in
line with data (within error bars) this points towards a possible deviation
from statistical equilibrium in the hadronization (at least for protons and
antiprotons).

Furthermore, we find that the $B {\bar B} \leftrightarrow 3 M$
reactions are more important at lower SPS or FAIR/NICA energies
where a net suppression for antiprotons and antihyperons up to a
factor of 2 -- 2.5 is seen in the PHSD calculations (cf. Fig.
\ref{9}). In this energy regime further data on antibaryons (also
with multiple strangeness) will be needed with high statistics as a
function of centrality and system size; a task well suited for the
upcoming facilities at FAIR and NICA. So far the baryon-antibaryon
dynamics is not sufficiently understood and open for further puzzles.


\begin{acknowledgments}
The authors acknowledge inspiring discussions with E. L. Bratkovskaya,
P. Moreau, A. Palmese, T. Steinert and R. Stock. We
thank the Helmholtz International Center for FAIR (HIC for FAIR),
the Helmholtz Graduate School for Hadron and Ion Research
(HGS-HIRe), and the Helmholtz Research School for Quark Matter Studies
in Heavy-Ion Collisions (H-QM) for support. The computational
resources have been provided by the Center for Scientific Computing
(CSC) in the framework of the Landes-Offensive zur Entwicklung
Wissenschaftlich-\"okonomischer Exzellenz (LOEWE) and the Green IT
Cube at FAIR.
\end{acknowledgments}


\begin{thebibliography}{99}
\bibitem{lQCD} Y. Aoki {\it et al.}, Phys. Lett. B {\bf 643}, 46 (2006).
\bibitem{lqcd0}  S. Borsanyi {\it et al.},
JHEP {\bf 1009}, 073 (2010); JHEP {\bf 1011}, 077 (2010); JHEP {\bf
1208}, 126 (2012).
\bibitem{LQCDx} S. Borsanyi {\it et al.}, Phys. Lett B {\bf 730}, 99
(2014); Phys. Rev. D {\bf 92}, 014505 (2015).

\bibitem{Peter} P. Petreczky [HotQCD Collaboration], PoS LATTICE
{ \bf 2012}, 069 (2012); AIP Conf. Proc. {\bf 1520}, 103 (2013).

\bibitem{Lat1} H.-T. Ding, F. Karsch, and S. Mukherjee, Int. J. Mod. Phys. E {\bf 24}, 1530007  (2015).

\bibitem{Lat2} A. Bazavov {\it et al.}, Phys. Rev. D {\bf 90},
094503 (2014).

\bibitem{CBMbook} P. Senger {\it  et al.}, Lect. Notes Phys. {\bf 814}, 681 (2011).

\bibitem{Zoltan17} A. Bazavov {\it et al.},
Phys. Rev. D {\bf 95}, 054504 (2017).

\bibitem{Karsch17} F. Karsch,
PoS CPOD2013, 046 (2013).

\bibitem{STM0}  F. Becattini and R.Fries, Landolt-Boernstein {\bf 23}, 208 (2010).

\bibitem{STM01} Braun-Munzinger, K.Redlich, and J.Stachel, in Quark-Gluon Plasma
3, eds. H.C.Hwa and X.N.Wang, World Scientific 2004, p.491.

\bibitem{STM02}  F. Becattini, M. Gadzdicki, A. Keranen, J. Manninen, and R. Stock,
Phys. Rev. C {\bf 69},024905 (2004).

\bibitem{STM03} A. Andronic {\it et al.}, Nucl. Phys. A {\bf 772}, 167 (2006).
\bibitem{STM04} J. Rafelski and M. Danos, Phys. Lett. B {\bf 97}, 167 (1980).
\bibitem{STM05} A. Tounsi and K. Redlich, J. Phys. G {\bf 28}, 2095 (2002).

\bibitem{STM1} P. Braun-Munzinger, V. Koch, T. Sch\"afer, and J. Stachel,
Phys. Rept. {\bf 621}, 76 (2016).

\bibitem{STM2} A. Andronic, D. Blaschke, P. Braun-Munzinger, J. Cleymans, K.
Fukushima {\it et al.},
Nucl. Phys. A {\bf 837}, 65 (2010).


\bibitem{PBM17}
A. Andronic, P. Braun-Munzinger, K. Redlich, and J. Stachel,
e-Print: arXiv:1710.09425


\bibitem{53} ALICE Collaboration, B. Abelev {\it et al.}, Phys. Rev.
C {\bf 88}, 044910 (2013).

\bibitem{Stock} F. Becattini {\it et al.}, Phys. Rev. C {\bf 85},
044921 (2012); Phys. Rev. Lett. {\bf 111}, 082302; Phys. Lett. B
{\bf 764}, 241 (2017).


\bibitem{62} ALICE Collaboration, J. Adam {\it et al.}, Nature Phys. {\bf 13}, 535 (2017).

\bibitem{Bass}
  S.~A.~Bass {\it et al.},
  Prog.\ Part.\ Nucl.\ Phys.\  {\bf 41}, 255 (1998).

\bibitem{Bleicher}
  M.~Bleicher {\it et al.},
  J.\ Phys.\ G {\bf 25}, 1859 (1999).

\bibitem{Francesco} F. Becattini, M. Bleicher, J. Steinheimer, and
R. Stock, arXiv:1712.03748.

\bibitem{Pratt}
Y. Pan and S. Pratt, Phys.Rev. C {\bf 89}, 044911 (2014).


\bibitem{Huovinen} P. Huovinen and J. I. Kapusta, Phys. Rev. C {\bf 69},
014902 (2004).

\bibitem{Cassing:2001ds}
  W.~Cassing,
  Nucl.\ Phys.\ A {\bf 700}, 618 (2002).

\bibitem{HSD}
  W. Cassing and E. L. Bratkovskaya, Phys. Rep. {\bf 308}, 65 (1999).
\bibitem{Weber}  H. Weber, E. L. Bratkovskaya,  { W. Cassing}, and H. St\"ocker,
Phys. Rev. C {\bf 67}, 014904 (2003).

\bibitem{Brat04} E. L. Bratkovskaya,
M. Bleicher, M. Reiter, S. Soff, H. St\"ocker, M. van Leeuwen,  S.
A. Bass, and W. Cassing,
Phys. Rev. C {\bf 69}, 054907 (2004).


\bibitem{Paper1} E. Seifert and W. Cassing, Phys. Rev. C {\bf 97},  024913 (2018).


\bibitem{PRC08}
  W. Cassing and E. L.  Bratkovskaya,  Phys. Rev. C {\bf 78}, 034919
  (2008).

\bibitem{PHSD}
  W. Cassing and E. Bratkovskaya, Nucl. Phys. A {\bf 831}, 215 (2009).

\bibitem{Bratkovskaya:2011wp}
  E.~L.~Bratkovskaya, W.~Cassing, V.~P.~Konchakovski and O.~Linnyk,
  Nucl.\ Phys.\ A {\bf 856}, 162 (2011).

\bibitem{Cas16} W. Cassing,  A. Palmese, P. Moreau,  and E. L.
Bratkovskaya, Phys. Rev. C {\bf 93}, 014902 (2016).

\bibitem{AlesPaper}
  A.~Palmese, W.~Cassing, E.~Seifert, T.~Steinert, P.~Moreau and E.~L.~Bratkovskaya,
  Phys.\ Rev.\ C {\bf 94},  044912 (2016).

\bibitem{review}  O. Linnyk, E. L. Bratkovskaya, and W. Cassing, Prog. Part. Nucl. Phys.
{\bf 87}, 50 (2016).

\bibitem{Linnyk:2015tha}
  O.~Linnyk {\it et al.},  Phys.\ Rev.\ C {\bf 88},  034904 (2013);
  Phys.\ Rev.\ C {\bf 92},  054914 (2015).


\bibitem{Konchakovski:2012yg}
  V.~P.~Konchakovski, E.~L.~Bratkovskaya, W.~Cassing, V.~D.~Toneev, S.~A.~Voloshin and V.~Voronyuk,
  Phys.\ Rev.\ C {\bf 85}, 044922 (2012).


\bibitem{Kadanoff-Baym}
  L.~P.~Kadanoff and G.~Baym,
  \textit{Quantum Statistical Mechanics}
  (Benjamin, New York, 1962).

\bibitem{Schwinger:1960qe}
  J.~S.~Schwinger,
  J.\ Math.\ Phys.\  {\bf 2}, 407 (1961).

\bibitem{Schwinger-Keldysh2}
  L.~V.~Keldysh,
  Sov.\ Phys.\ JETP {\bf 20}, 1018 (1965).

\bibitem{Botermans:1990qi}
  W.~Botermans and R.~Malfliet,
  Phys.\ Rept.\  {\bf 198}, 115 (1990).

\bibitem{Cassing:1999wx}
  W.~Cassing and S.~Juchem,
  Nucl.\ Phys.\ A {\bf 665}, 377 (2000);  Nucl.\ Phys.\ A {\bf 672}, 417 (2000).




\bibitem{Peshier:2004bv}
  A.~Peshier,
  Phys.\ Rev.\ D {\bf 70}, 034016 (2004).

\bibitem{Peshier:2004ya}
  A.~Peshier,
  J.\ Phys.\ G {\bf 31}, S371 (2005).

\bibitem{NilssonAlmqvist:1986rx}
  B.~Nilsson-Almqvist and E.~Stenlund,
  Comput.\ Phys.\ Commun.\  {\bf 43}, 387 (1987).


\bibitem{Ex1}
  S.~S.~Adler {\it et al.} [PHENIX Collaboration],
  Phys.\ Rev.\ C {\bf 69}, 034909 (2004).


\bibitem{Ex4}
  K.~Aamodt {\it et al.} [ALICE Collaboration],
  Phys.\ Rev.\ Lett.\  {\bf 106}, 032301 (2011).

\bibitem{Ex5}
  B.~Abelev {\it et al.} [ALICE Collaboration],
  Phys.\ Lett.\ B {\bf 720}, 52 (2013);
  Phys.\ Rev.\ C {\bf 88}, 044910 (2013);
  Phys.\ Rev.\ Lett.\  {\bf 111}, 222301 (2013);
  Phys.\ Lett.\ B {\bf 728}, 216 (2014).

\bibitem{Ex6}
  E.~Abbas {\it et al.} [ALICE Collaboration],
  Phys.\ Lett.\ B {\bf 726}, 610 (2013);
  Phys.\ Rev.\ Lett.\  {\bf 111}, 222301 (2013).

\bibitem{Ex7}
  J.~Adam {\it et al.} [ALICE Collaboration],
  Phys.\ Lett.\ B {\bf 754}, 373 (2016).


\bibitem{Koncha2015} V. P. Konchakovski, { W. Cassing}, and V.D.
Toneev,
J. Phys. G {\bf 42}, 055106 (2015).


\bibitem{Ex2}
  G.~Agakishiev {\it et al.} [STAR Collaboration],
  Phys.\ Rev.\ Lett.\  {\bf 108}, 072301 (2012).

\bibitem{Ex3}
  J.~Adams {\it et al.} [STAR Collaboration],
  Phys.\ Rev.\ Lett.\  {\bf 98}, 062301 (2007).


\bibitem{Ex8}
  K.~Adcox {\it et al.} [PHENIX Collaboration],
  Phys.\ Rev.\ Lett.\  {\bf 89}, 092302 (2002);
  Phys.\ Rev.\ C {\bf 69}, 024904 (2004).

\bibitem{Ex9}
  C.~Adler {\it et al.} [STAR Collaboration],
  Phys.\ Rev.\ Lett.\  {\bf 89}, 092301 (2002).

\bibitem{Ex10}
  J.~Adams {\it et al.} [STAR Collaboration],
  Phys.\ Rev.\ Lett.\  {\bf 92}, 182301 (2004).

\bibitem{Ex11}
  L.~Adamczyk {\it et al.} [STAR Collaboration],
  Phys.\ Rev.\ C {\bf 96}, no. 4, 044904 (2017).

\bibitem{Ex12}
  L.~Ahle {\it et al.} [E802 Collaboration],
  Phys.\ Rev.\ Lett.\  {\bf 81}, 2650 (1998).

\bibitem{Ex13}
  L.~Ahle {\it et al.} [E866 and E917 Collaborations],
  Phys.\ Lett.\ B {\bf 476}, 1 (2000);
  Phys.\ Lett.\ B {\bf 490}, 53 (2000).

\bibitem{Ex14}
  B.~B.~Back {\it et al.} [E917 Collaboration],
  Phys.\ Rev.\ Lett.\  {\bf 87}, 242301 (2001).

\bibitem{Ex15}
  J.~L.~Klay {\it et al.} [E-0895 Collaboration],
  Phys.\ Rev.\ C {\bf 68}, 054905 (2003).

\bibitem{Ex16}
  P.~Chung {\it et al.} [E895 Collaboration],
  Phys.\ Rev.\ Lett.\  {\bf 91}, 202301 (2003).


\bibitem{Ex17}
  S.~V.~Afanasiev {\it et al.} [NA49 Collaboration],
  Phys.\ Rev.\ C {\bf 66}, 054902 (2002).

\bibitem{Ex18}
  C.~Alt {\it et al.} [NA49 Collaboration],
  Phys.\ Rev.\ Lett.\  {\bf 94}, 192301 (2005);
  Phys.\ Rev.\ C {\bf 73}, 044910 (2006);
  Phys.\ Rev.\ C {\bf 77}, 024903 (2008);
  Phys.\ Rev.\ C {\bf 78}, 034918 (2008).

\bibitem{Ex19}
  T.~Anticic {\it et al.} [NA49 Collaboration],
  Phys.\ Rev.\ Lett.\  {\bf 93}, 022302 (2004);
  Phys.\ Rev.\ C {\bf 80}, 034906 (2009);
  Phys.\ Rev.\ C {\bf 83}, 014901 (2011).

\bibitem{Ex20}
  M.~K.~Mitrovski {\it et al.} [NA49 Collaboration],
  J.\ Phys.\ G {\bf 32}, S43 (2006).
\end{thebibliography}
\end{document}